\documentclass[10pt,preprint]{aastex}

\begin{document}

\title{Neutral Hydrogen in the Interacting Magellanic Spirals NGC 4618/4625}

\author{Stephanie J. Bush}
\affil{Departments of Astronomy and Physics, Case Western Reserve University, 10900 Euclid Ave., Cleveland, OH 44106}
\email{sjb16@cwru.edu}

\and

\author{Eric M. Wilcots}
\affil{Astronomy Department, University of Wisconsin, 
475 N. Charter St., Madison, WI 53706}
\email {ewilcots@astro.wisc.edu}

\begin{abstract}

Asymmetry is a common trait in spiral galaxies and is particularly frequent among Magellanic spirals. To explore how
morphological and kinematic asymmetry are affected by companion
galaxies, we analyze neutral hydrogen observations of the interacting
Magellanic spirals NGC 4618 and 4625.  The analysis of the H~I distribution revealed that about 10\% of the total H~I
mass of NGC 4618 resides in a looping tidal structure that appears to
wrap all the way around the galaxy. Through calculations based on
derived H~I profiles, we show that NGC 4618 and 4625 are no more
asymmetric than non-interacting Magellanic spirals analyzed by Wilcots
\& Prescott (2004). We also derive rotation curves for the approaching
and receding sides of each galaxy. By fitting the mean curves with an
isothermal halo model, we calculate dynamical masses of $4.7 \times
10^{9}$ M$_{\odot}$ and $9.8 \times 10^{9}$ M$_{\odot}$ out to
6.7 kpc, for NGC 4618 and 4625 respectively. While the
rotation curves had systematically higher velocities on the receding
side of each galaxy, the effect was no more pronounced than in studies
of non-interacting spirals (Swaters et. al. 1999). The degree of
interaction-driven asymmetry in both galaxies is indistinguishable from
the intrinsic degree of asymmetry of lopsided galaxies.

\end{abstract}


\section{Introduction}

Barred Magellanic spirals represent an intermediate phase in the Hubble
Sequence between late-type spiral galaxies and irregular
galaxies. They are characterized by a single, strong, spiral arm; a
bright, off-center bar; and often a high degree of asymmetry in
morphology and/or kinematics (de Vaucouleurs \& Freeman 1972). While
asymmetry is very pronounced in Magellanic spirals, it is also a
common characteristic of a large percentage of all disk galaxies (Richter
\& Sancisi 1994), and therefore important to our understanding of disk
galaxies' structure in general.

The frequency of strong asymmetry in spirals has led to a number of
different theories to explain its origin. Based on the N-body
simulations of Walker, Mihos \& Hernquist (1996), Zaritsky \& Rix
(1997) proposed that minor mergers could induce strong asymmetry in
spirals. Their theory is supported by photographic surveys which found 
a high incidence of optical companions in Magellanic spirals
(Odewahn 1994). However, these models suggest a transient asymmetry,
lasting only one or two orbits (Walker, Mihos \& Hernquist
1996). Subsequent analytical studies by Sparke and collaborators
(Levine \& Sparke 1998; Noordermeer, Sparke \& Levine 2001) indicate
that the one-armed morphology of Magellanic spirals could be quite
long-lived if the galaxies' disks are offset from the dynamical center
of the halo. To investigate these theories, Wilcots \& Prescott (2004)
completed a H~I survey of 13 Magellanic spirals from the Odewahn
(1994) survey. However, they only confirmed 4 interacting
systems. This suggests that current interactions are not responsible for the
lopsidedness of Magellanic spirals and that the phenomenon must be longer
lived as in the models of Noordermeer, Sparke \& Levine (2001). Though
the role of interactions in shaping Magellanic spirals is uncertain we
cannot escape the observation that many of the prototypical
Magellanics (e.g. NGC 4618, the LMC) are part of binary (if not
triple) systems. Odewahn (1994) suggested that Magellanic spirals with
the highest arm strength also had the most obvious optical
companions. However, the effect of binarity on asymmetry is not
understood. Here we address whether apparently binary systems are
interacting and what effect these interactions may or may not be
having on the participating galaxies' asymmetry and kinematics.

To explore interaction effects and asymmetry, sufficiently
detailed studies of galaxies' kinematics are needed.  For example, the
model rotation curves presented by Noordermeer, Sparke, \& Levine 
(2001) reflect a disparity between the approaching and receding
sides of the galaxy.  Additionally, a
detailed analysis of the galaxies' rotation curves allows us to
determine the degree to which the halo dominates the dynamics of the
galaxy. H~I distributions are often more extended than stellar
distributions and therefore allow us to derive rotation curves farther
into the galactic halo. Extended H~I distributions are also better
tracers of gravitational interactions. However, sufficiently high
resolution H~I data only exists for a handful of Magellanic spirals
(e.g. Swaters et al. 1999).

We present high resolution neutral hydrogen data for the interacting
Magellanic spiral pair NGC 4618/4625. Both galaxies show the optical
characteristics of the class mentioned above: a single, strong spiral
arm emanating from a bar that appears to be offset from the center of
the galaxy (Odewahn 1991). While an earlier H~I study of these
galaxies did not find evidence of an interaction (van
Moorsel 1983), our more sensitive data may. From our data we were
able to analyze asymmetry and dynamics by producing H~I profiles,
velocity fields and rotation curves. These galaxies turn out to be remarkably symmetric, despite their close proximity in the viewing plane. Our observations are explained in
\S 2 and the data is presented in \S 3. In \S 4.1 we explore the dynamics of the galaxies, we discuss the nature of the H~I loop in \S 4.2, and we calculate interaction timescales and degrees of morphological asymmetry in \S 4.3 and \S 4.4, respectively. We summarize our results in \S 5.  

\section{Observations and Data Reduction}

Our data was taken using the Very Large Array (VLA){\footnote{The VLA
is part of the National Radio Astronomy Observatory which is operated
by Associated Universities Inc., under a cooperative agreement with
the National Science Foundation.}} in the C configuration. The
integrated time spent on NGC 4618/4625 was $\sim$ 400 minutes in
$\sim$ 50 minute sessions, pausing after each session to observe the
phase calibrator, 1225+368, for $\sim$ 3-4 minutes. The amplitude and
bandpass calibrator 1328+307 was observed twice for $\sim$ 10 minutes
each time. We used the AIPS tasks {\it vlacalib} and {\it vlaclcal} to
calibrate the amplitude and phase. The task {\it imagr} was used to
create and CLEAN a naturally weighted image cube. Three thousand CLEAN
iterations were needed to fully remove the side lobes. Note that a
bowl remains in our CLEAN cube. This is the result of the absence of
short spacing data and the presence of extended low column density
gas.  Several line free channels on either end of the data cube were
fit and subtracted off the entire cube using the AIPS task {\it
uvlin}. The original cube covered a velocity range of 255.9 km s$^{-1}$ to
850.4 km s$^{-1}$, with a velocity resolution of 5.2 km s$^{-1}$. The final
synthesized beam size was $19.6^{\prime\prime} \times 16.9
^{\prime\prime}$ and our final $1\sigma$ sensitivity was $2 \times
10^{18}$ cm$^{-2}$ per channel.

\section {Results}
 
Our main results are presented in the moment maps in Figures 1 and
2, which have been smoothed by a factor of 2. These maps also have an applied flux cut of $4\sigma$ ($8 \times
10^{18}$ cm$^{-2}$) in each channel. Along with the integrated H~I map, Figure 1 shows H~I contours
overlaid on an optical image from the Digitized Sky Survey.  NGC 4618
is in the southwest at a position of (12$^{h}$41$^{m}$35$^{s}$
41$^{\circ}$08$^{\prime}$23$^{\prime\prime}$) and shows an elongated
HI distribution with an H~I ring nearly encircling the disk. NGC 4625 is in the northeast at (12$^{h}$41$^{m}$52$^{s}$
41$^{\circ}$16$^{\prime}$18$^{\prime\prime}$) and shows a very large
H~I disk with a tidal feature extending off the northwest corner. The
most striking features of the H~I distributions are their large
extents and the H~I ring encircling NGC 4618. We discuss the nature of this ring in \S4.2. We derived radii of the galaxies by calculating the radius of a
circle with the area covered by each galaxy. The results were
235.9$^{\prime\prime}$ for NGC 4618 and 235.5$^{\prime\prime}$ for NGC
4625 down to $3\sigma$ ($3.54 \times 10^{20}$ cm$^{-2}$).  Adopting
effective optical radii of 1.5 kpc and 0.7 kpc, as well as a distance
of 6.0 Mpc from Odewahn (1991), we get a ratio of H~I extent to
effective optical radii of $\sim$ 4.6 for NGC 4618 and $\sim$ 9.8 for
NGC 4625. NGC 4625 has one of the largest H~I/optical ratios of any
galaxy yet mapped, comparable to the extent of DDO 154 (Carignan \&
Purton 1998) and larger than NGC 4449 (Hunter et. al. 1998).

The integrated flux was used to calculate total H~I masses of $(5.3 \pm 0.6)
\times 10^{8}$ M$_{\odot}$ and $(3.9 \pm 0.6) \times 10^{8}$
M$_{\odot}$ for NGC 4618 and 4625 respectively. Since it was difficult
to separate the lower intensity features from the noise at $3\sigma$,
these masses were calculated to $4\sigma$ ($4.72 \times 10^{20}$ cm$^{-2}$). These are comparable to van Moorsel's H~I masses of $4.4 \times 10^{8}$ M$_{\odot}$ and $2.8 \times 10^{8}$ M$_{\odot}$ (1983). Our slightly increased mass is indicative of our better sensitivity to low column density material such as the H~I ring around NGC 4618 and the tidal material extending off NGC 4625 which van Moorsel did not detect. We also calculated the mass
in the H~I ring on NGC 4618 using $5\sigma$ contours to separate the ring from the disk. The east side contains $(4.1\pm 0.6)\times
10^{7}$ M$_{\odot}$ of H~I and the west side contains $(5.7\pm0.6)\times
10^{7}$ M$_{\odot}$.

The velocity field is presented in Figure 2. NGC 4625 shows characteristic
differential rotation while NGC 4618 has a very disturbed velocity
field. We created position-velocity diagrams by rotating the data cube so
that the major axis was coincident with one of the principle axes of
the cube. These are shown in Figures 3 and 4. Figure 4 shows the P-V
diagram for NGC 4625 along the major axis. It is typical of a disk
galaxy, with a linear rise in circular velocity at small radii
followed by a flat rotation curve. For NGC 4618, P-V diagrams along
both the major and minor axes are shown (Figure 3). The diagram along
the major axis shows the dynamics of the galaxy disk, exhibiting an
almost solid body rotation throughout much of the disk, but with
a sharp turnover at large radii that is indicative of the prominent warp
evident in the velocity in Figure 2. The P-V diagram along the minor axis
gives a clearer impression of the dynamics of the H~I ring and will
be discussed in \S 4.2. 

\section{Analysis and Discussion} 

\subsection{Rotation Curves and Mass Modeling}

In order to quantify the dynamics and possible kinematic asymmetry of
the galaxies, we fit a tilted ring model to the approaching and
receding sides of each galaxies' observed velocity field (Figure 2)
using the AIPS task {\it gal}. In the models the systemic velocity of
the galaxies was held constant at the mean values calculated in \S
4.4. (The FWZM value for NGC 4618 was not included in the mean used
for NGC 4618 because of its strong dependence on the H~I ring.) The
position of the center of the galaxy was allowed to vary but stayed
extremely consistent. The position angle of the major axis and
inclination of the disk were allowed to vary and are plotted with the
fitted rotation curves in Figures 5 and 6. Note that since the outer
portion of NGC 4618 is very perturbed by the interaction, the tilted
ring model only fit accurately out to 200$^{\prime\prime}$, 85\% of
the observed radii. The rotational velocities are consistent with
other Magellanic and late type spirals. They are comparable to van Moorsel's (1983) previous work, with differences of around 10-20 km s$^{-1}$ that are most likely due to the van Moorsel's (1983) constant correction for disk inclination.Van Moorsel (1983) adopted a constant inclination of 35$^{\circ}$ for NGC 4618 and 27$^{\circ}$ for NGC 4625, while our fitted inclinations for NGC 4618 range between  40$^{\circ}$ and  80$^{\circ}$ and between 20$^{\circ}$ and  40$^{\circ}$ for NGC 4625.

It is interesting that both
galaxies show an increased maximum velocity on their receding side
compared to their approaching side. In NGC 4618, this divergence appears
around $100^{\prime\prime}$, roughly coincident with the edge of the
optical disk, and grows as we progress to large radii. The position
angle and inclination diverge at about the same point, most likely
because at this radii we are analyzing disturbed H~I. To the limits of
our resolution, NGC 4625 also shows the velocities diverging at the
edge of the optical disk. Beyond the optical disk it shows a
consistent velocity divergence of about 10 km s$^{-1}$. The position angle
and inclination of either side of NGC 4625 stay roughly consistent
with one another, although the position angles show a slow increasing
trend. The mean rotation curve of both galaxies flattens at large
radii.

To model the dynamical mass we fit an isothermal halo model (De Blok
et. al. 2001, Carignan \& Purton 1998) to the mean of the approaching
and receding sides of the tilted ring models for each galaxy. The
isothermal halo has the density profile:
\begin{equation}
\rho_{iso}(R)=\rho_{0}\left[1+\left(\frac{R}{R_{C}}\right)^{2}\right]^{-1}
\end{equation}
where $\rho_{0}$ is the central density of the halo and $R_{C}$ is the
halo core radius. This creates the rotation curve:
\begin{equation}
V(R)=\sqrt{4\pi G \rho_{0}R_{C}^{2}\left[1-\frac{R_{C}}{R}\arctan \left( \frac{R}{R_{C}}\right)\right]}
\end{equation}
The fitted parameters are $V_{\infty}$ and $R_{C}$ which relate to the
parameters in equations (1) and (2) through:
\begin{equation}
V_{\infty} =\sqrt{4\pi G \rho_{0}R_{C}^{2}}
\end{equation}
We calculate the dynamical mass by numerically integrating equation 2
out to 6.7 kpc. Our fitted parameters were $V_{\infty}=65
\pm 15$ km s$^{-1}$ and $R_{C}=1.45 \pm 0.33$ kpc for NGC 4618 and
$V_{\infty}=83.7 \pm 0.4$ km s$^{-1}$ and $R_{C}=0.41 \pm 0.03 $ kpc for NGC
4625. These values gave central densities of 0.037 M$_{\odot}$ pc$^{-3}$
and 0.77 M$_{\odot}$ pc$^{-3}$, respectively.  Our fits are shown in
Figures 7 and 8. While the fit for NGC 4618 is well within the 
error bars of the observed rotation curve, the model overestimates the circular velocity of the outer disk of NGC 4625.  The mass calculated
for NGC 4618 was $4.7 \times 10^{9}$ M$_{\odot}$. Adopting B-band stellar luminosities from Odewahn (1991)
and assuming a mass to light ratio of 1 M$_{\odot}$/L$_{\odot}$, we
calculate stellar masses of $2.37 \times 10^{9}$ M$_{\odot}$ and $0.42
\times 10^{9}$ M$_{\odot}$, for NGC 4618 and 4625, respectively. When combined with the our calculated H~I masses, this gives NGC 4618 a disk mass
fraction of 0.6. The fitted mass for NGC 4625 was $9.8 \times 10^{9}$
M$_{\odot}$ for a disk mass fraction of 0.08, but since the disk mass to light ratio could be considerably larger than 1 M$_{\odot}$/L$_{\odot}$ this is probably a 
lower limit. These masses are within
the expected range for Magellanic spirals (Wilcots \& Prescott 2004)
and the disk mass fraction for NGC 4625 is also reasonable. However,
the disk mass fraction for NGC 4618 is much higher than we would
expect. It has been shown that interactions can strongly affect
rotation curves (Barton et. al. 1999) so it is possible that the
interaction is affecting the disk in such a way that we see a slowed
or truncated rotation curve and calculate a low halo mass.

\subsection{Nature of the H~I Loop}

We explore two different interpretations of the H~I ring surrounding NGC 4618. 
The velocity field and position-velocity diagram 
of the tidal material surrounding NGC
4618 is indicative of a single tidal arm that fell back towards the
galaxy and wrapped around it, forming a coherent loop. The formation
of such loops is common in simulations of major mergers
(e.g. Hernquist \& Spergel 1992, Hibbard \& Mihos 1995). The
kinematics of the tidal material are apparent in the position velocity
diagram of NGC 4618 along its minor axis (Figure 3). NGC 4618's minor
axis is effectively the major axis of the loop. In Figure 3, the tidal
material's dynamics show striking similarity to a solid body rotation
curve, supporting the assertion that it is a coherent feature.

Alternately, the loop could be interpreted as a galactic warp possibly created by the interaction, where what appears to be a gap between the galaxy and the ring is actually filled with very low column density gas. If the H~I ring is a galactic warp, we can fit a tilted ring model to the H~I ring. In order to do this, we spatially smoothed the data cubes by a factor of two, and reanalyzed the data. The derived tilted ring model is Figure 9. The tilted ring model was fit to either side of the galaxy out to 120$^{\prime\prime}$, but was fit better to the entire galaxy beyond 120$^{\prime\prime}$. The rotation curve for the outer portions of the galaxy appears to follow well from the rotation curve from the inner portions, with the exception of three points at radii of 270, 290 and 310$^{\prime\prime}$ that have significantly lower velocities than all other points. Data at these radii lie in the 'gap' between the disk and the ring and is not an accurate portrayal of the rotation curve. The apparent continuation of the rotation curve in the outer ring shown would support the interpretation of the ring as a galactic warp. The outer disk has a substantially different position angle than the inner disk, changing from around 180$^{\circ}$ in the inner disk to around 120$^{\circ}$ in the outer ring. However it has retained roughly the same inclination, 60$^{\circ}$, as the unperturbed disk. This indicates that this structure could be a continuation of the disk, but it does not appear to be a warp in the traditional sense of a change in inclination in the outer portions of the disk (Garcia-Ruiz et. al. 2002). Though it is possible that the H~I ring is simply a continuation of the disk, the mass distribution this would imply for NGC 4618 is very unusual, and we believe the tidal loop interpretation is a more coherent picture of the evolution of the galaxy.

However, If the ring is a continuation of the disk, we can use this model to derive a more appropriate mass for NGC 4618. The isothermal halo fit to the data is in Figure 10 (note that the three points in the 'gap' were neglected). The derived fitting parameters were $V_{\infty}=73\pm 5$ km s$^{-1}$ and $R_{C}=1.8 \pm 0.2$ kpc, which gave a central density of 0.03 M$_{\odot}$ pc$^{-3}$, a new mass of $1.1 \times 10^{10}$ M$_{\odot}$  and a disk mass fraction of 0.26. Although this result is more reasonable, it is still significantly larger than the disk mass fraction for NGC 4625.

\subsection{Interaction Timescales}

Using our observations we can make a few basic arguments about the
timescale of the galaxies' interaction. By dividing the projected
distance between the centers of the galaxies (14.5 kpc) by their
projected velocity difference roughly de-projected by a factor of $\sqrt{2}$, we
calculate the minimum time since the galaxies have interacted to be
$\sim$ 0.14 Gyrs. However, this assumes the galaxies are on a
completely radial orbit across the line of sight, so the true time to travel even the projected
distance is actually several factors larger.

By analyzing the dynamics of H~I loop encircling NGC 4618 as a tidal loop, we can estimate the interaction timescale
another way.  The
smooth nature of this loop indicates that the galaxies have only had
one close passage, had there been additional close passages we would
not expect the shell to survive as a coherent structure (Hernquist \&
Spergel 1992). By adopting the tidal materiel's projected major axis
as the radius of its orbit and its velocity at that position as the
orbital velocity, we calculate an orbital timescale of $\sim$ 0.5-0.7
Gyrs. This is on the same order as each galaxies' orbital time and
significantly larger than the estimate derived above. In addition, it
would take the tidal material an additional fraction of an orbit time
to leave the disk and enter orbit. 

\subsection{Asymmetry}

The degree of asymmetry for each galaxy can be quantified several ways using H~I
profiles which are presented in Figure 11. We follow the treatment of
Wilcots \& Prescott (2004) and measure asymmetry in two ways (Haynes
et. al. 1998). First, we calculate the systemic velocity of the
galaxies by calculating the full width of the profile at zero
intensity, 20\% and 50\% of the peak. For a completely symmetric
galaxy these three values should be equal. Second, we integrated under the
approaching and receding sides of the peak and compared the resulting
areas. Once again, for a symmetric galaxy they should be equal. The
results are in Table 1.  Note that, despite its disturbed appearance, NGC 4618's areal ratio was very close to one, and its FWHM and FW20 are
very similar. The FWZM has been strongly affected by asymmetry between
the sides of the H~I ring and therefore is not a good indicator of the symmetry
of the disk. In contrast, NGC 4625 has an areal ratio of 1.3 and a
standard deviation of 3.6 km s$^{-1}$ in the systemic velocities.

The striking thing about this symmetry analysis is how
morphologically symmetric these galaxies really are, given that they
are interacting. We argue above that the interaction between NGC 4618
and 4625 has been ongoing for at least 0.5 Gyr, yet it appears to be significantly affecting only the asymmetry of the outermost
gas.  In a survey of 13 Magellanic spirals, Wilcots \& Prescott
(2004) found that the mean areal asymmetry was the same for Magellanic
spirals with companions as it was for galaxies without companions
(both were 1.1). Similarly, the standard deviation of the median
velocities was 9.3 km s$^{-1}$ for galaxies with companions and 7.4 km s$^{-1}$ for
galaxies without companions. The H~I profile of the {\it disk} of NGC
4618 is actually {\it more} symmetric than the average for Magellanic
spirals in Wilcots \& Prescott (2004).

The interaction is likely to be affecting the galaxies' kinematic
asymmetry as well. Binary galaxies often show asymmetric rotation
curves, including examples similar to NGC 4618 and 4625, where one
side of the rotation curve falls off while the other continues to rise
(Barton et. al. 1999). Each galaxy shows a systematically faster
rotation curve on its receding side and it is possible that this is
due to disk warping caused by the interaction. However, it is very
difficult to determine whether this is truly an interaction effect or if
it is an intrinsic trait of the galaxies, as suggested by Swaters
et. al.  (1999). These authors examined the H~I velocity
fields of DDO 9 and NGC 4365 and found rotation curves that flattened
on one side but continued to rise on the other. However, these
galaxies were {\it not} observably interacting, so their kinematic
asymmetry is most likely intrinsic. From this, we can draw the
conclusion that even for galaxies known to be interacting, the
possible effects of the interaction, at least in the early stages, are
indistinguishable from the intrinsic asymmetry seen in significant
numbers of galaxies that are not interacting.

This data reveals that the encounter between NGC 4618 and
NGC 4625 appears to be significantly influencing only the outermost H~I,
which is beyond the optical extent of the galaxy.  The impact of
the interaction on the inner H~I disk, and by extrapolation the stellar
disk, appears to be minimal.  Yet, both galaxies have strongly asymmetric
optical morphologies (Odewahn 1991).  The inference one draws is that the
current interaction cannot be responsible for the observed degree of
asymmetry in the stellar disk, unless the response of the disks to the
encounter is incredibly rapid.

\section{Conclusions}

We have shown that NGC 4618 and NGC 4625 are
interacting. The majority of the material exterior to the disks in the system forms a tidal loop encircling NGC 4618 and amounts to $\sim 10\%$ of the total
H~I mass. One of our most interesting observations is that although
the outermost gas on NGC 4618 has been strongly affected by the
interaction, NGC 4625 and the inner
parts of NGC 4618 seem mostly unperturbed. It is also interesting to note
that the velocity fields appear to be unperturbed out to the optical
radius of each galaxy, at which point the receding side of the galaxy
has a consistently higher velocity. This suggests that the optical
component stabilizes the H~I disk.

An important factor in explaining the appearance of these
galaxies may be the disk to halo mass ratio. We derived total masses
of $4.7 \times 10^{9}$ M$_{\odot}$ and $9.8 \times 10^{9}$ M$_{\odot}$
for NGC 4618 and NGC 4625 respectively. When combined with measures of
the optical luminosity (Odewahn 1991) and our measures of the total
H~I mass, we find NGC 4625 has a $M_{disk}$/$M_{halo}$ ratio of only
8\% while NGC 4618's disk accounts for almost 60\% of its total
mass. Prescott et. al. (2004) use NGC 4618's stellar velocity
dispersion to find a disk mass of $2 \times 10^{9}$ M$_{\odot}$,
giving a ratio of 40\%. Alternately, if the outermost gas on NGC 4618 is interpreted as a warp, we find a total mass of $1.1\times 10^{10}$M$_{\odot}$, giving a disk mass fraction of 26\% or 18\% In either case, NGC 4618 appears to have a
rather massive disk with respect to the halo. Either NGC 4618
actually does have a particularly low mass to light ratio, or the outer 
disk or the loss or warping of the outer portions of NGC 4618, have lead to an 
underestimate of the total dynamical mass.

It is clear that the interaction involves two galaxies with
significantly different $M_{disk}$/$M_{halo}$ ratios. Gerin
et. al. (1990) indicate that an encounter between a galaxy with a
massive halo and a total mass 1.5 times the mass of the target will
result in rapid bar formation in the target galaxy but very slow
development in the galaxy with the massive halo. We may be witnessing
an interaction that has had the effect of enhancing an already
lopsided structure in NGC 4618 while leaving the halo-dominated NGC
4625 relatively unperturbed. The level of disturbance in the morphology of each galaxy's H~I distribution may also be explained by how far into the halo the stellar disks extend.  Dubinski et. al. (1996) used numerical simulations of  merging galaxies to explore how 
tidal tail strength varies with halo mass and extent and found that galaxies whose halo causes a deeper potential well create weaker tidal tails. If NGC 4625 is embedded in a large extended halo, the halo's deep potential well could have prevented the disk from warping or creating large tidal features.

This work demonstrates the inherent difficulty in understanding the
origin of the strong asymmetries characteristic of Magellanic
spirals. Our analysis of the asymmetry of NGC 4618 and 4625's H~I
profiles and velocity fields show no more asymmetry than a sample of
non-interacting Magellanic spirals (Wilcots \& Prescott 2004). Thus,
H~I asymmetry is not a reliable indicator of either the presence of a
companion or the effects of an interaction. The asymmetry between the
approaching and receding side of the rotation curves are comparable to
what Barton et. al. (1999) suggest for binary galaxies as well as to
simulations of lopsided galaxies presented by Noordermeer
et. al. (2001). Again, this shows how difficult it is to find 
compelling evidence that
the interaction is the cause of the lopsidedness in NGC 4618 and
NGC 4625. High resolution observations of other binary systems and a more
thorough theoretical examination of the impact of close passages on
the structure and kinematics of galaxies with a range of halo
strengths may continue to shed light on this problem.

\section{Acknowledgments}

We would like to thank Dr. Stacy S. McGaugh for allowing us to use his Fortran code to fit dynamical masses and  Dr. J. Christopher Mihos for useful discussions about galaxy interactions. This work was supported by a NSF-REU site grant (AST-0139463) to the University of Wisconsin-Madison. EMW was
supported by a NSF grant AST-9875008. This research has made use of the NASA/IPAC Extragalactic
Database (NED) which is operated by the Jet Propulsion Laboratory,
California Institute of Technology, under contract with the National
Aeronautics and Space Administration. The Digitized Sky Survey was
produced at the Space Telescope Science Institute under
U.S. Government grant NAG W-2166. The images of these surveys are
based on photographic data obtained using the Oschin Schmidt Telescope
on Palomar Mountain and the UK Schmidt Telescope. The plates were
processed into the present compressed digital form with the permission
of these institutions.

\clearpage

\begin{deluxetable}{ccccccc}
\tabletypesize{\scriptsize}
\tablecaption{H~I Profile Asymmetries \label{tbl-4}}
\tablewidth{0pt}
\tablehead{
\colhead{Galaxy} & \colhead{FWZM} & \colhead{FWHM} & \colhead{FW20} & \colhead{Mean} &
\colhead{$\sigma$} & \colhead{Areal Ratio} \\
\colhead{} & \colhead{km s$^{-1}$} & \colhead{km s$^{-1}$} & \colhead{km s$^{-1}$} & \colhead{km s$^{-1}$} & \colhead{km s$^{-1}$} & \colhead{}  }
\startdata
NGC 4618 & 544.7 & 533.0 & 534.8 & 537.5 &  5.1 & 1.005 \\
NGC 4625 & 599.0 & 605.8 & 607.4 & 604.1 & 3.6 & 1.29 \\
\enddata
\end{deluxetable}
\clearpage

\begin{figure}
\plottwo{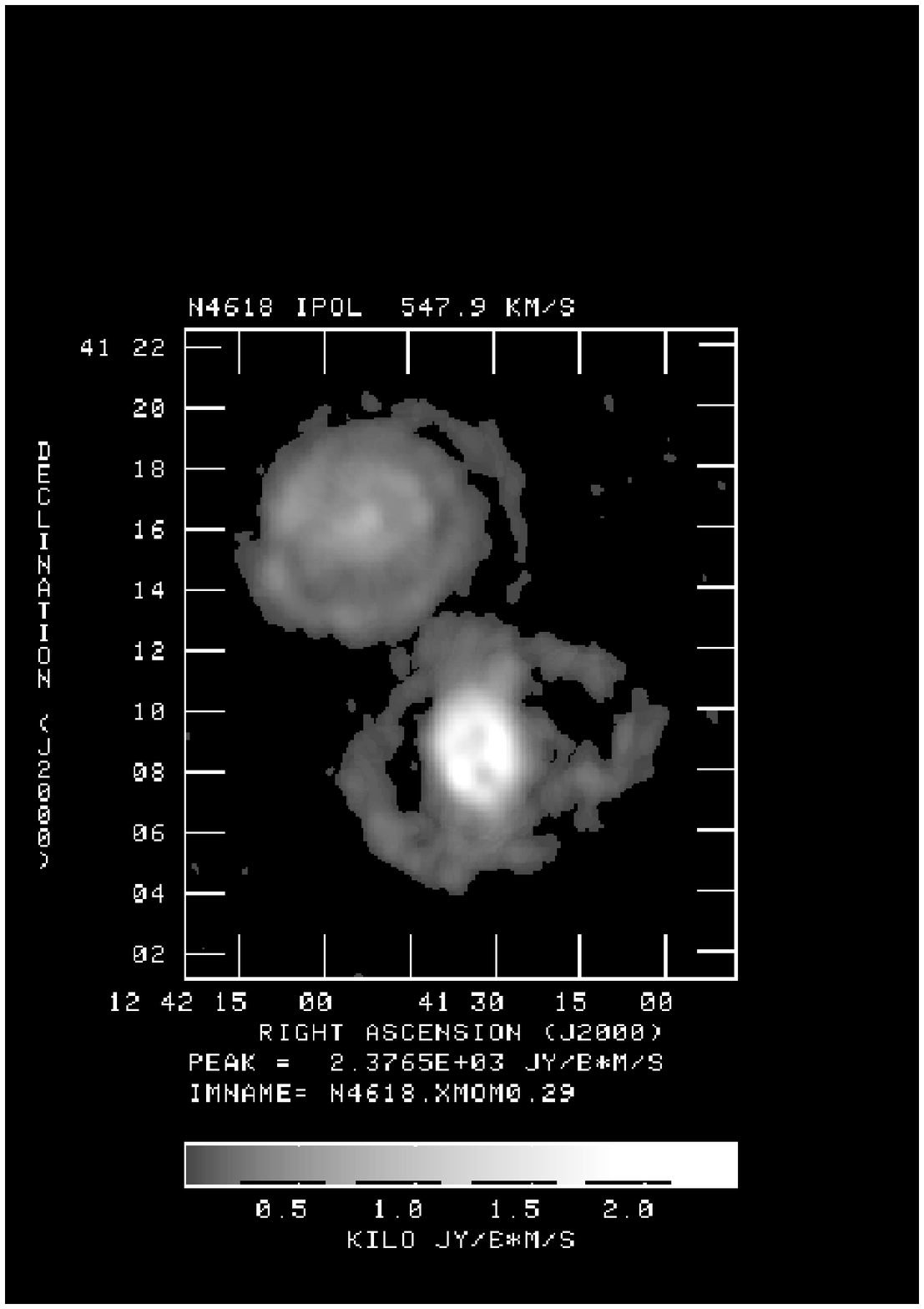}{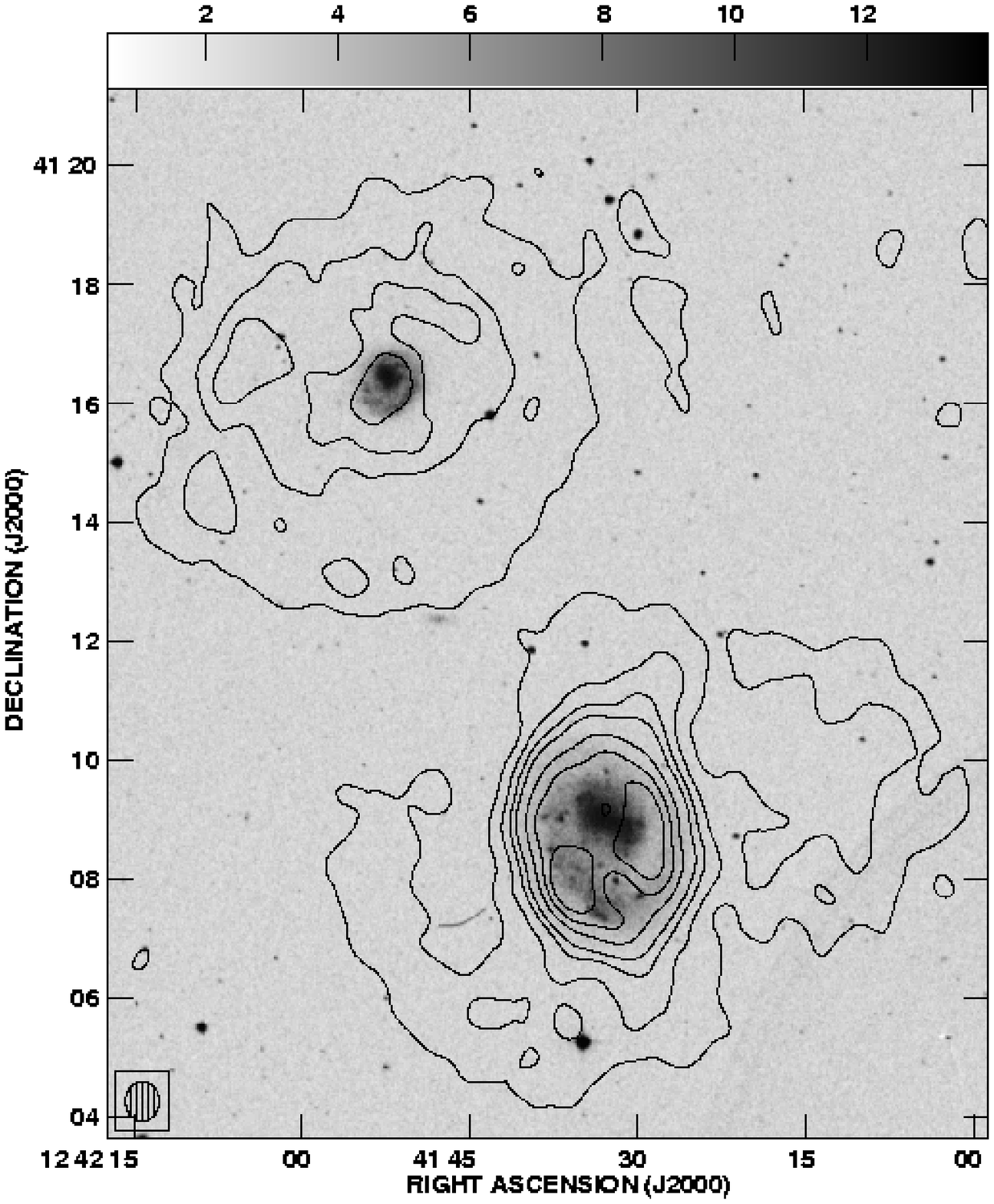}
\caption{ The integrated H~I map in gray-scale (left) and H~I contours overlaid on an optical image from the Digitized Sky Survey (DSS). For clarity, these plots have been smoothed by a factor of 2. H~I contours are placed at 5, 10, 15, 20, 30, 40 and 50 $\times (1.18\times10^{20}$ cm$^{-2}$). The beam is plotted in the lower left corner of the contoured image.\label{fig1}}
\end{figure}

\begin{figure}
\plottwo{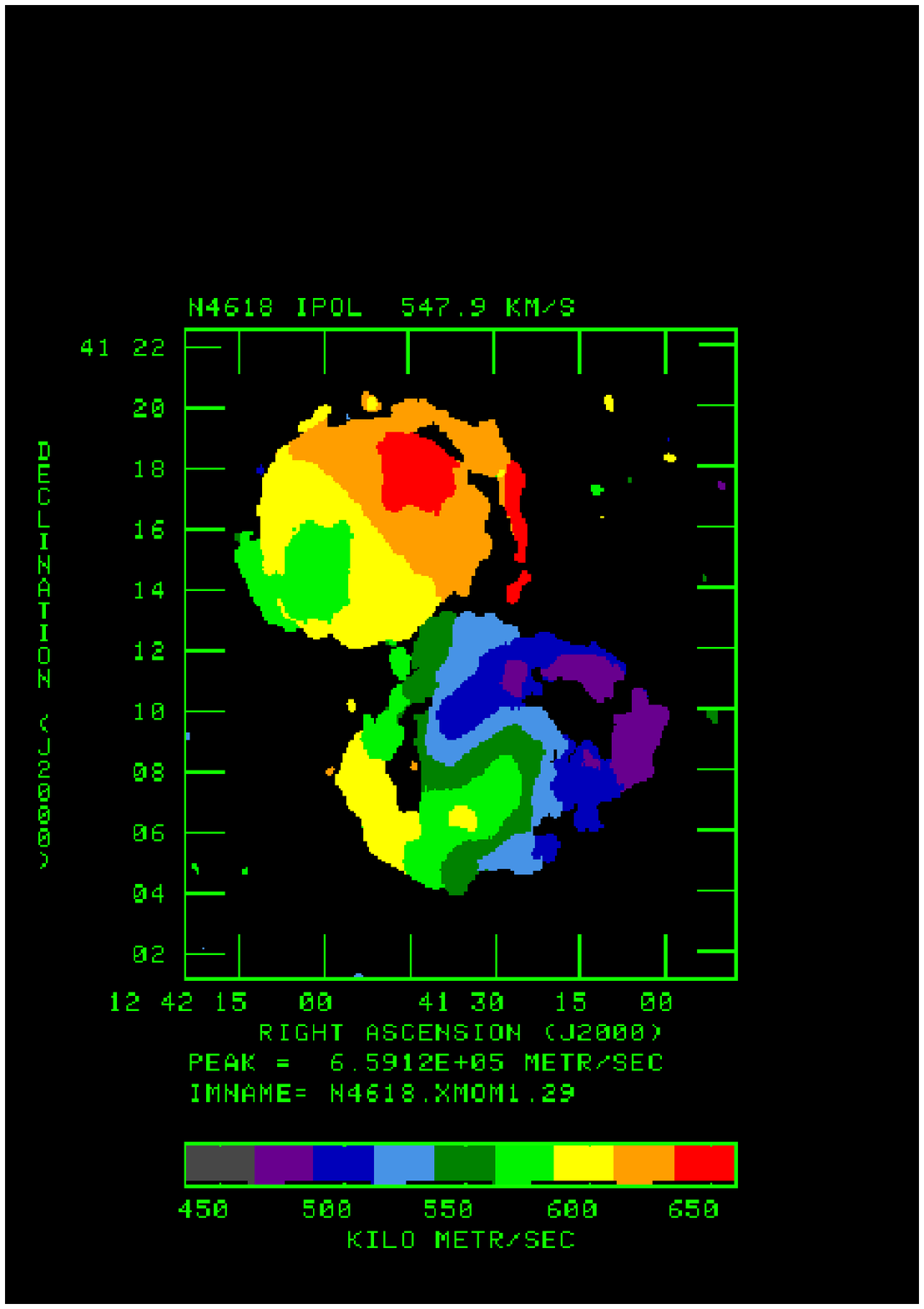}{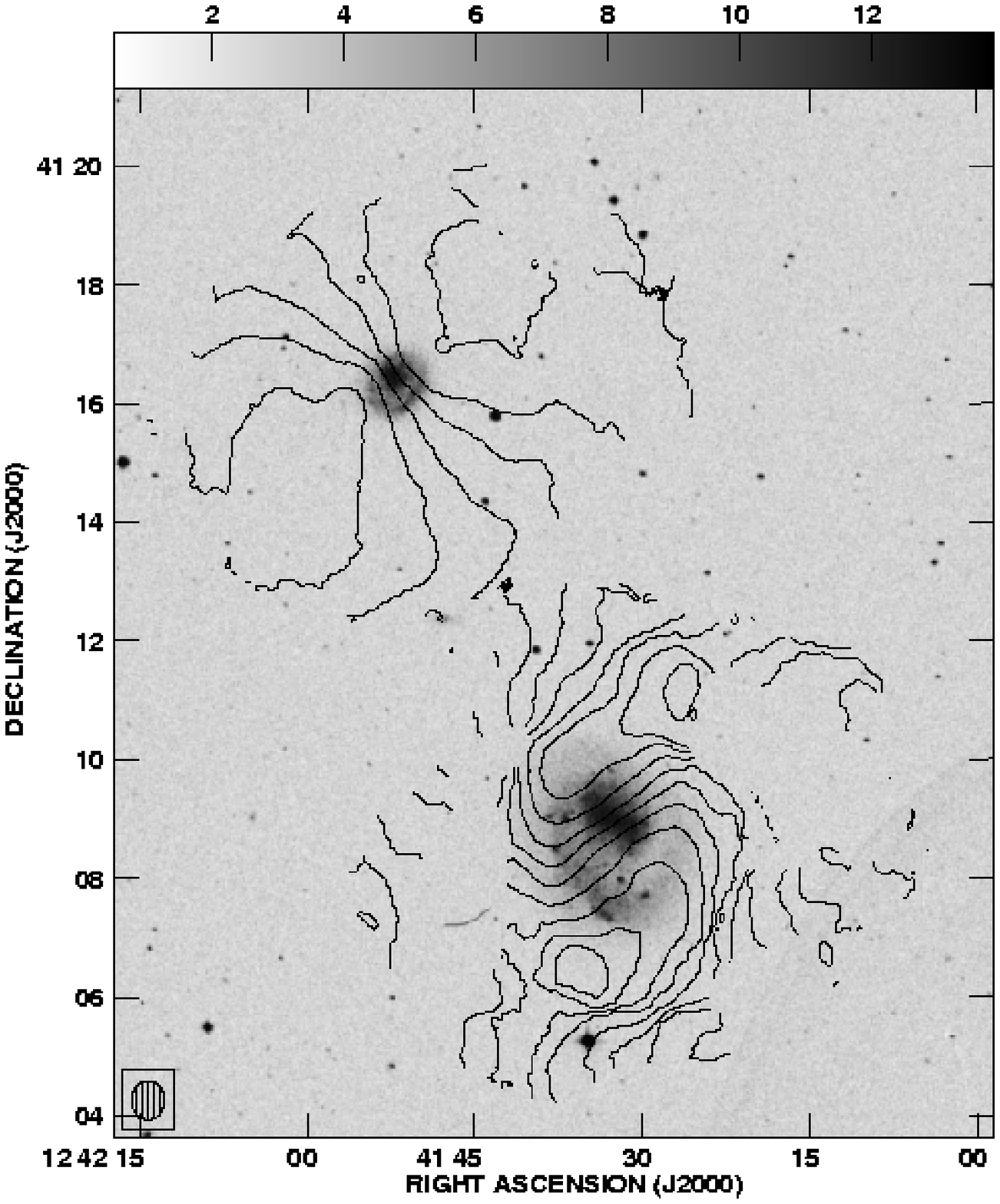}
\caption{The color-coded velocity field (left) and isovelocity contours overlaid on the DSS optical image (right). For clarity, these plots have been smoothed by a factor of 2. Contours are spaced by 10 km s$^{-1}$. The contours on NGC 4618 increment from 480 km s$^{-1}$ in the northwest to 580 km s$^{-1}$ in the southeast and the contours on NGC 4625 increment from 580 km s$^{-1}$ in the southeast to 630 km s$^{-1}$ in the northwest. Note that the contours on NGC 4618 appear to change abruptly at the edge of the optical disk.  The beam is plotted in the lower left corner of the contoured image. \label{fig2}}
\end{figure}

\begin{figure}
\plottwo{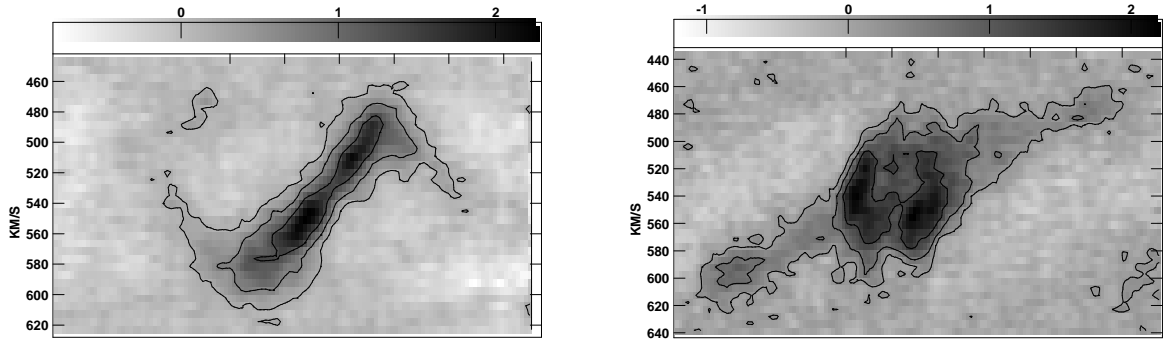}{bush.fig3b.ps}
\caption{Contoured position velocity diagrams for NGC 4618 along the major axis (left) and minor axis (right). The contours are simply used to highlight structure and are at 3, 10 and 20 times the noise in the gray-scale.  \label{fig3}}
\end{figure}

\begin{figure}
\plotone{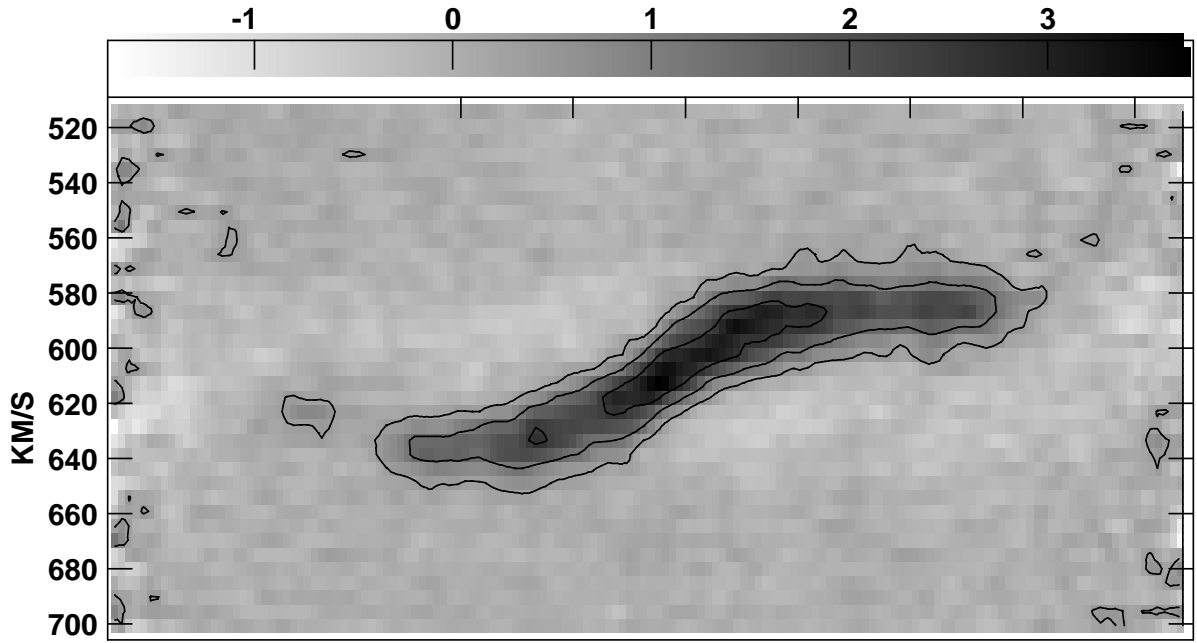}
\caption{Contoured position velocity diagram for 4625. Contours are used simply to highlight structure and are at 3, 10 and 20 times the noise in the gray-scale.  \label{fig4}}
\end{figure}

\begin{figure}
\plotone{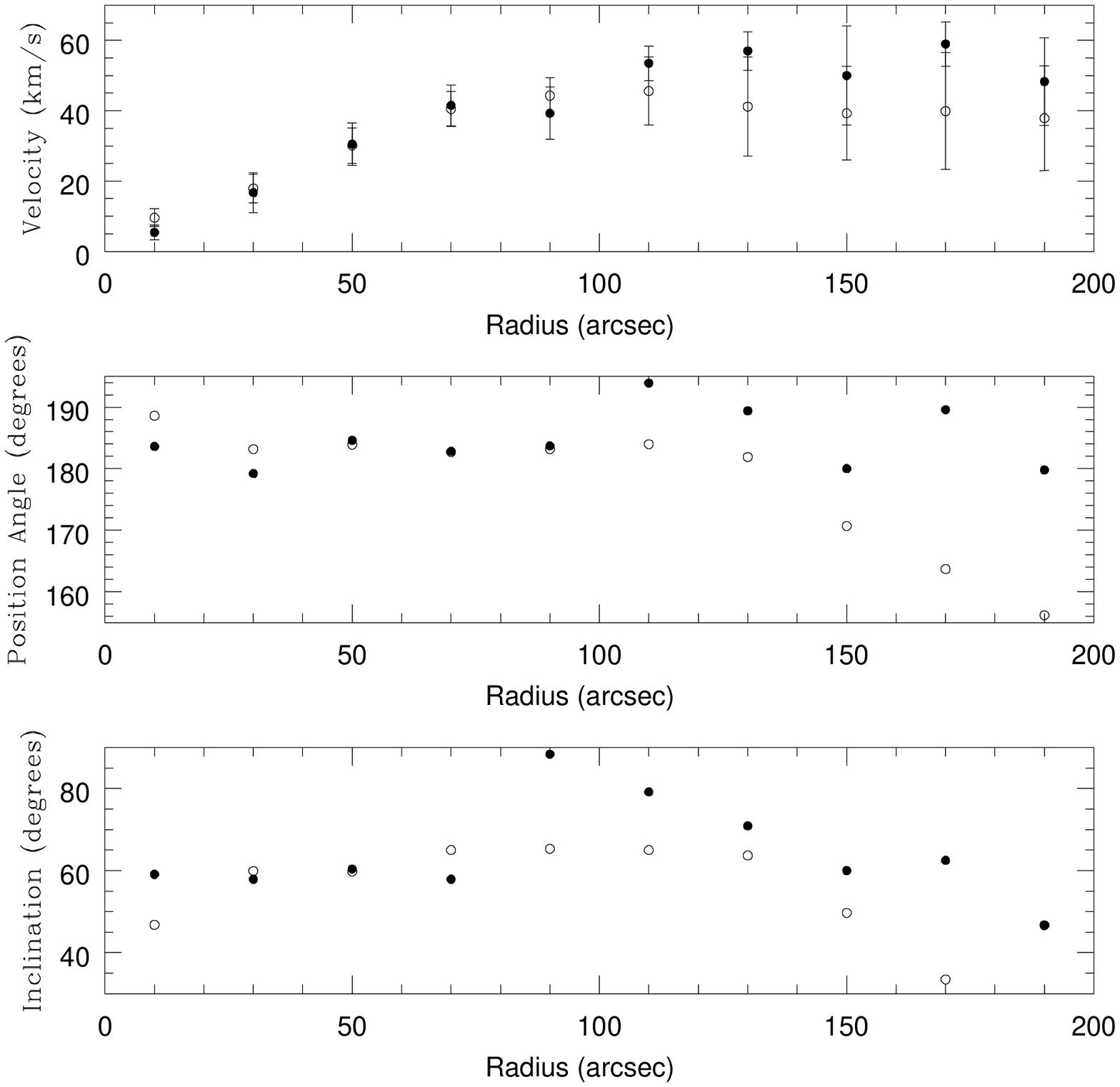}
\caption{Tilted ring model for NGC 4618 with 20$^{\prime\prime}$ rings. Open circles represent the approaching side and closed circles represent the receding side. Error estimates were calculated by AIPS based on the number of points used to fit the model.  \label{fig6}}
\end{figure}

\begin{figure}
\plotone{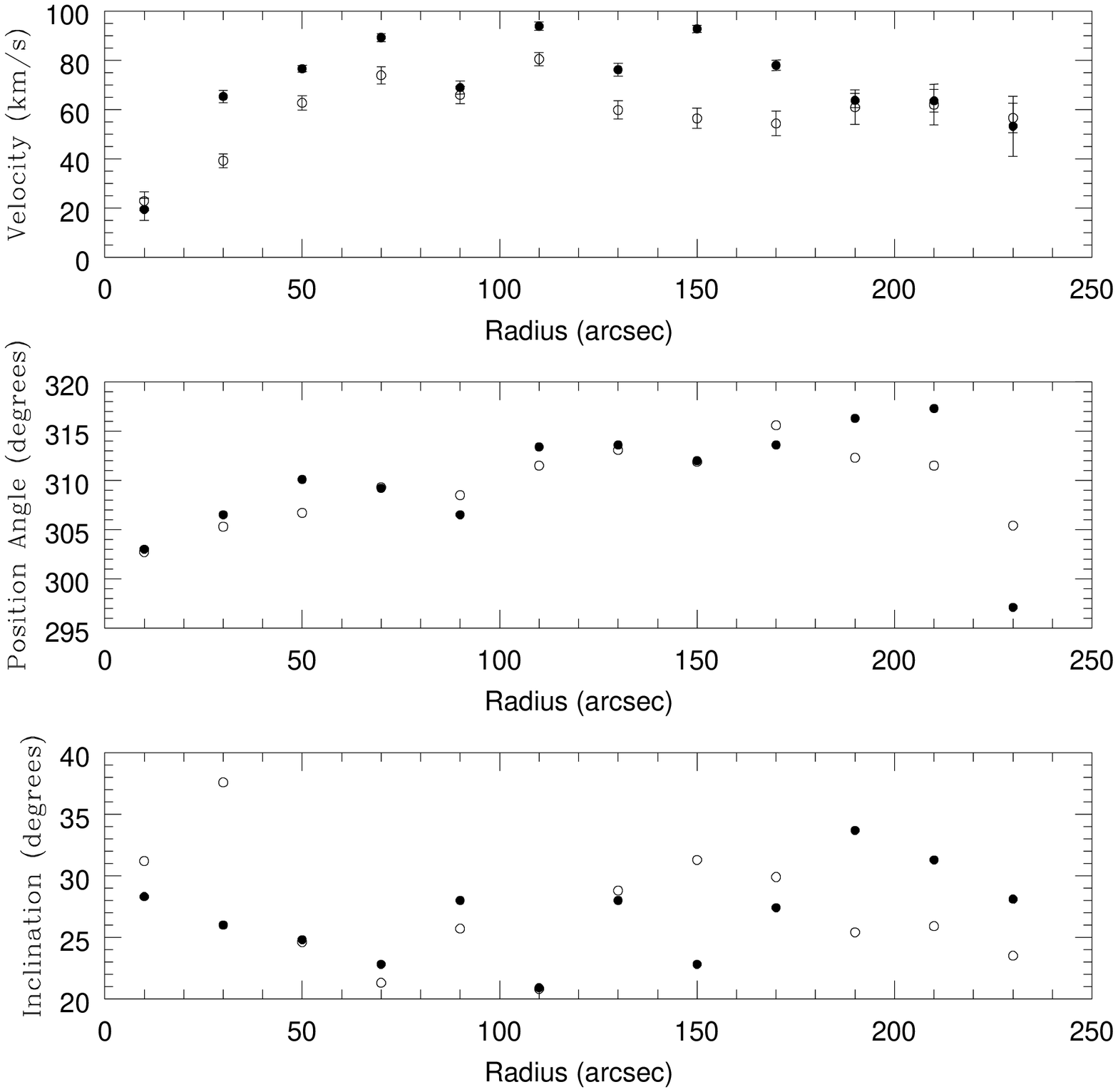}
\caption{Tilted ring model for NGC 4625 with 20$^{\prime\prime}$ rings. Open circles represent the approaching side and closed circles represent the receding side. Error estimates were calculated by AIPS based on the number of points used to fit the model.    \label{fig7}}
\end{figure}

\begin{figure}
\plotone{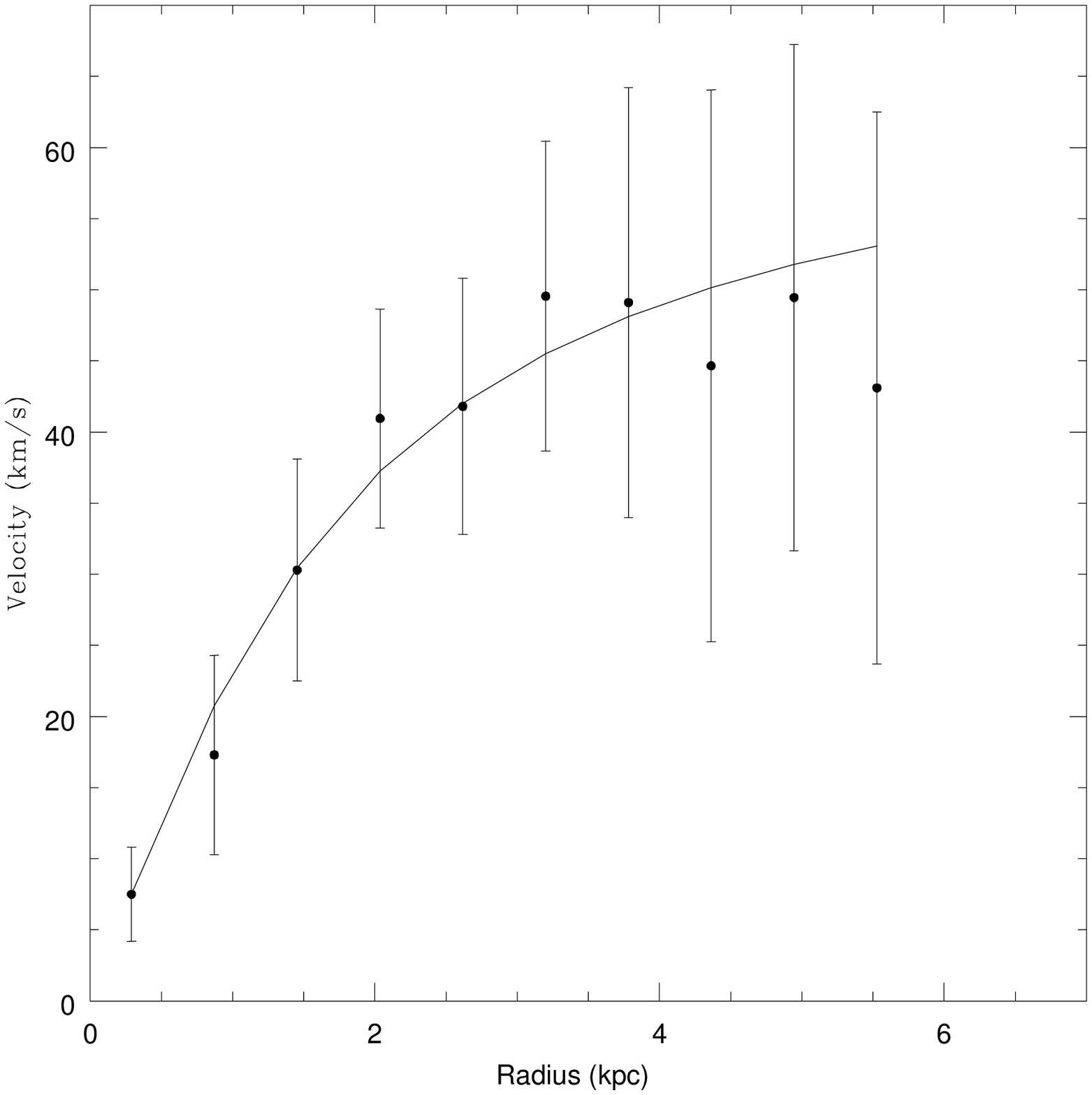}
\caption{Isothermal halo model fit to the tilted ring model for NGC 4618. The radii are computed assuming a distance of 6 Mpc (Odewahn 1991). \label{fig8}}
\end{figure}

\begin{figure}
\plotone{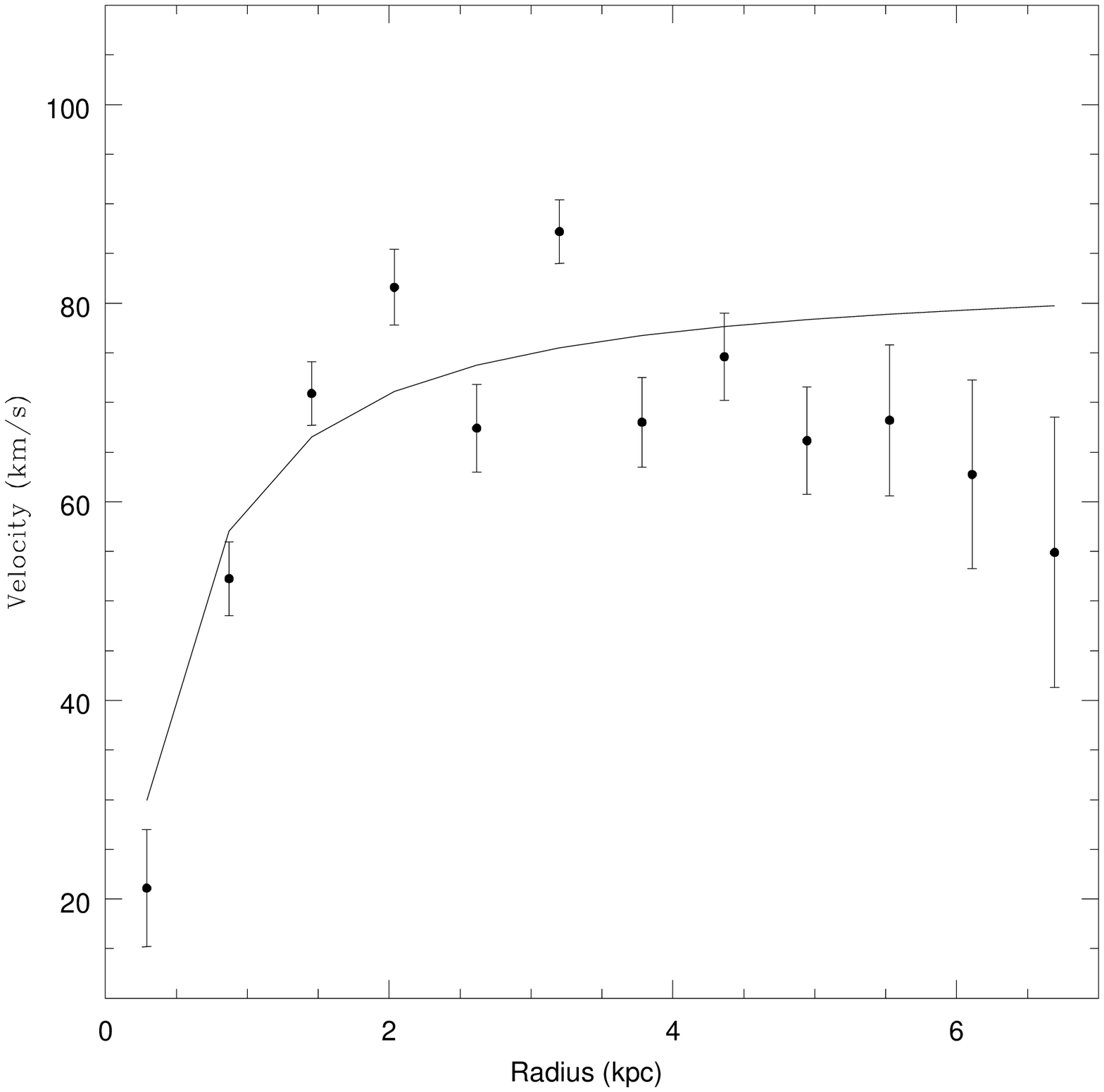}
\caption{Isothermal halo model fit to the tilted ring model for NGC 4625. The radii are computed assuming a distance of 6 Mpc (Odewahn 1991). \label{fig9}}
\end{figure}

\begin{figure}
\plotone{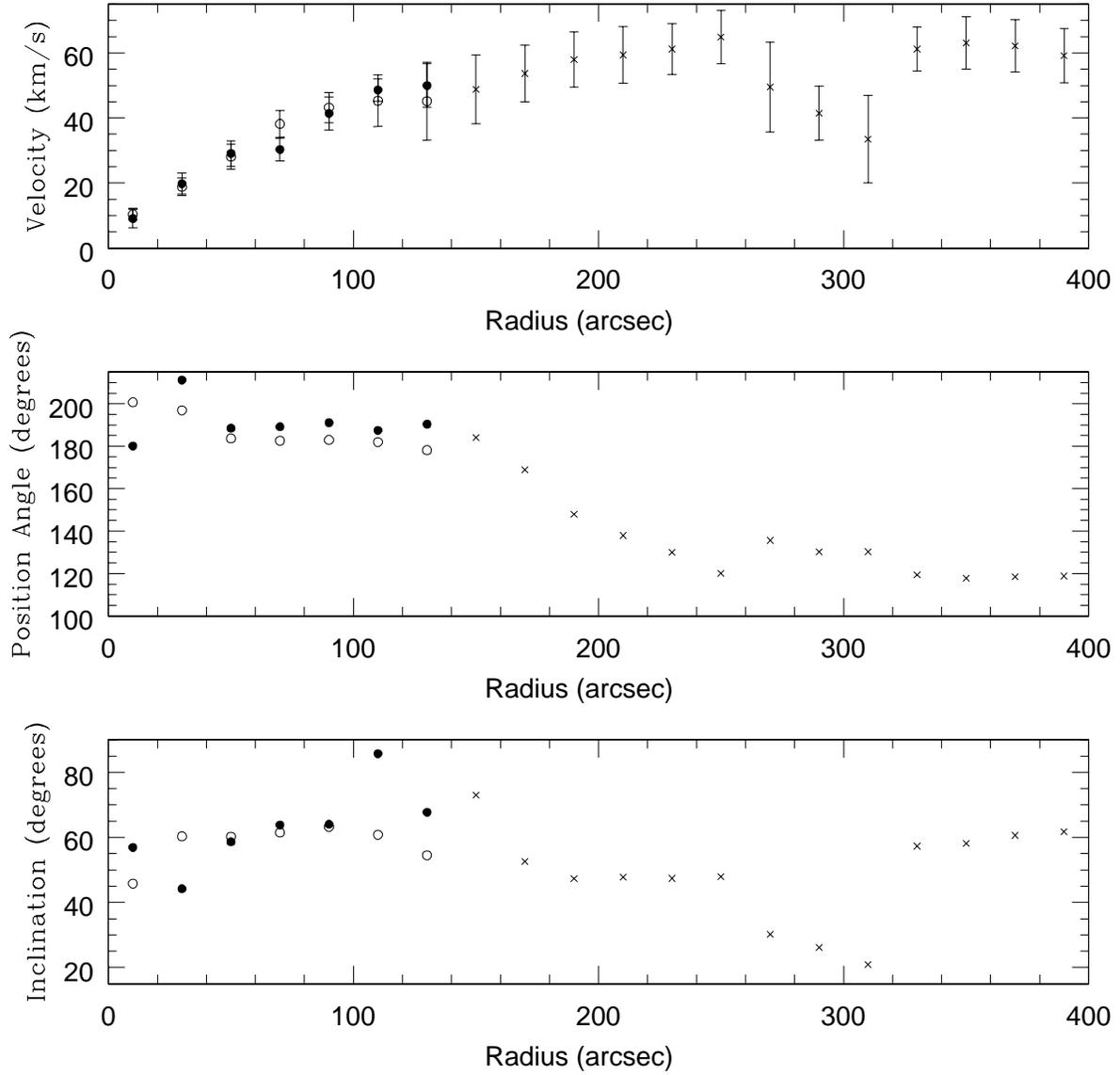}
\caption{Tilted ring model for NGC 4618 with the smoothed data and 20$^{\prime\prime}$ rings. Open circles represent the approaching side and closed circles represent the receding side. Crosses represent the fit in the outer regions. Error estimates were calculated by AIPS based on the number of points used to fit the model.\label{fig10}}
\end{figure}

\begin{figure}
\plotone{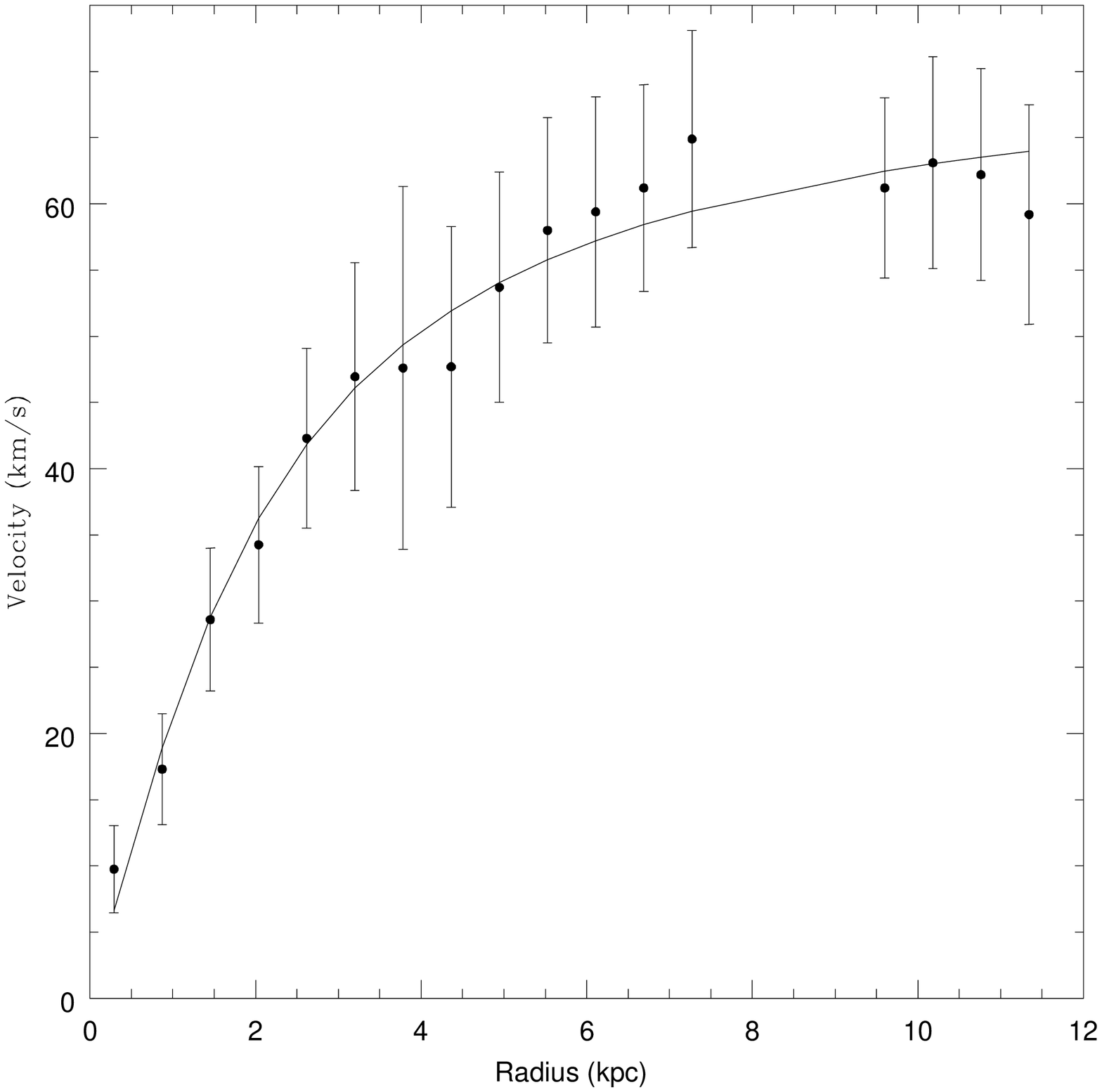}
\caption{Isothermal halo model fit to the tilted ring model for NGC 4618 smoothed data. The radii are computed assuming a distance of 6 Mpc (Odewahn 1991). \label{fig11}}
\end{figure}

\begin{figure}
\plotone{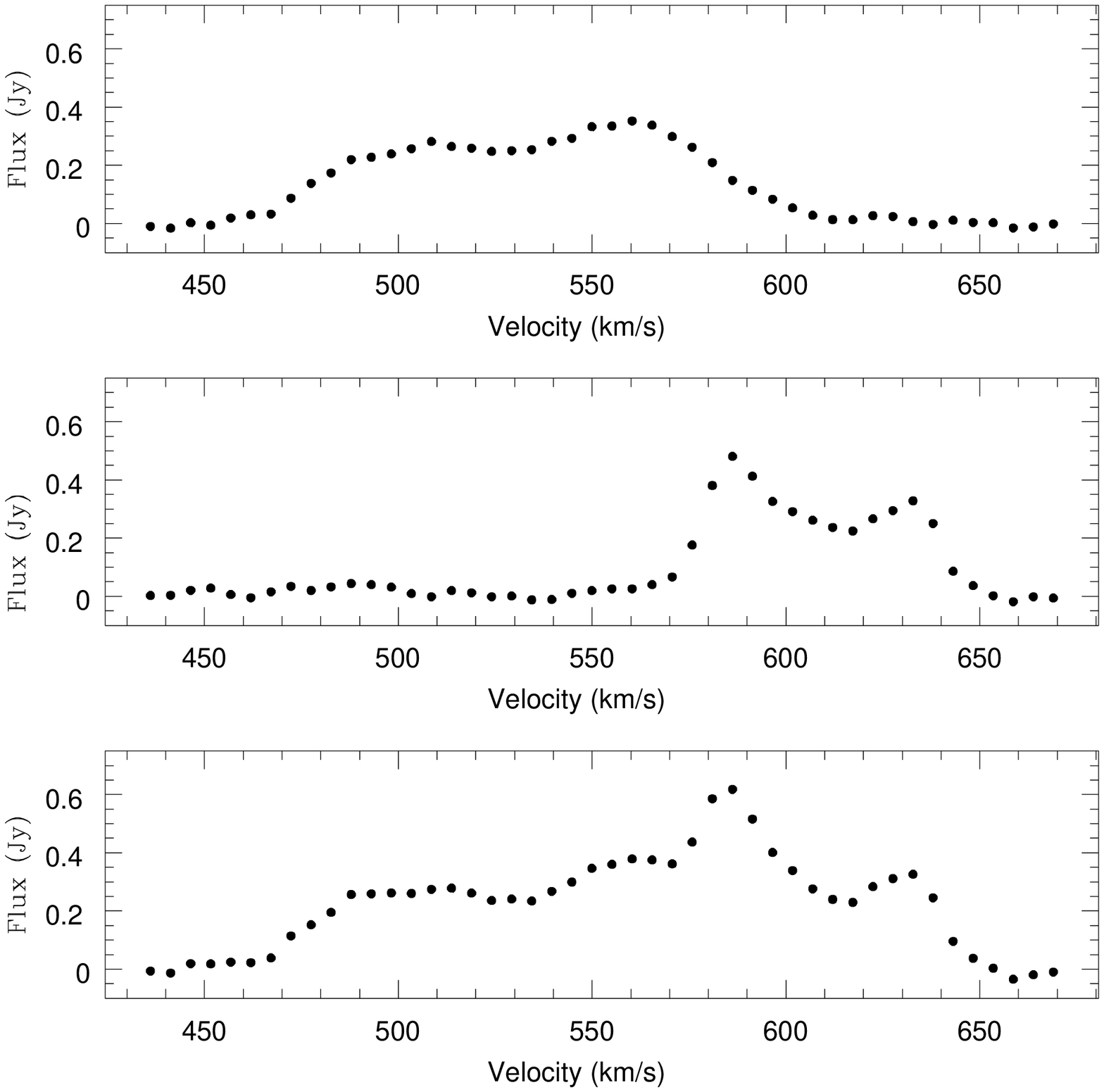}
\caption{H~I intensity profiles for NGC 4618 (top), NGC 4625 (middle) and the entire system (bottom).   \label{fig5}}
\end{figure}


\clearpage

\begin{references}

\reference{} Barton, E.J., Bromley, B.C. \& Geller, M.J. 1999, ApJ, 511, L25.
\reference{} Carignan, C. \& Purton, C. 1998, ApJ, 506, 125.
\reference{} de Blok, W.J.G., McGaugh, S.S. \& Rubin, V.C. 2001, AJ, 122, 2396.
\reference{} de Vaucouleurs, G. \& Freeman, K.C. 1972, Vistas in Astronomy, 14, 163.
\reference{} Dubinski, J., Mihos, J.C., \& Hernquist, L. 1996, ApJ, 462, 576.
\reference{} Garc{\'{\i}}a-Ruiz, I., Sancisi, R., \& Kuijken, K.\ 2002, \aap, 394, 769. 
\reference{} Gerin, M., Combes, F., \& Athanassoula, E. 1990, A\&A, 230, 37.
\reference{} Haynes, M.P., Hogg, D.E., Maddalena, R.J., Roberts, M.S., \& van Zee, L. 1998, AJ, 115, 62. 
\reference{} Hernquist, L. \& Spergel, D. 1992, ApJ, 399, L117.
\reference{} Hibbard, J.~E.~\& Mihos, J.~C.\ 1995, \aj, 110, 140. 
\reference{} Hunter, D.A., Wilcots, E.M., van Woerden, H., Gallagher, J.S. \& Kohle, S. 1998, ApJ, 495, L47.
\reference{} Levine, S.E. \& Sparke, L. S. 1998, ApJ, 496, L13. 
\reference{} Noordermeer, E., Sparke, L.S., \& Levine, S.E. 2001, MNRAS, 328, 1064.
\reference{} Odewahn, S.C. 1991, AJ, 101, 829.
\reference{} Odewahn, S.C. 1994, AJ, 107, 1320.
\reference{} Prescott, M.K., Wilcots, E.M., Bershady, M.A., \& Westfall, K. 2004, in preparation.
\reference{} Richter, O.-G. \& Sancisi, R. 1994, A\&A, 290, L9.
\reference{} Swaters, R.A., Schoenmakers, R.H.M, Sancisi, R. \& van Albada, T.S. 1999, 305, 330.
\reference{} van Moorsel, G.A. 1983, A\&AS, 54, 19.
\reference{} Walker, I.R., Mihos, J.C. \& Hernquist, L. 1996, ApJ, 460, 121.
\reference{} Wilcots, E.M., \& Prescott, M.K.M. 2004, AJ, 127, 1900.
\reference{} Zaritsky, D. \& Rix, H-W. 1997, ApJ, 477, 118.

\end{references}
\end{document}